\documentclass[aps,twocolumn,showpacs,floatfix,10pt]{revtex4-1}

\usepackage[dvips]{graphicx}
\usepackage{amsmath}
\usepackage{amssymb}
\usepackage[nice]{nicefrac}
\usepackage{color}

\begin{document}

\title{Density fluctuations and random walks in an overdamped and supercooled simple liquid}

\author{E. B. Postnikov}
\affiliation{Department of Theoretical Physics, Kursk State University, Radishcheva st., 33, 305000 Kursk, Russia}
\email{postnicov@gmail.com}

\begin{abstract}
In this work, the short-time dynamics of simple liquid is explored both analytically and numerically with the focus on the interplay between the density fluctuations in a volume surrounding a chosen particle and its random walk motion. The particles interact via the Lennard-Jones potential with parameters corresponding to liquid argon.  For large times, analytical calculations based on the fluctuation theory provides an explicit expression reproducing isothermal change of the self-diffusion coefficient in liquid argon corresponding to the experimental data. These results lead to the conclusion that such behavior is based on the  reduced mobility of particles reflected in their density fluctuations that can be equivalently achieved in the cases of either low temperatures and pressures (supercooling) or moderate temperatures and high pressures (overdamping). 
\end{abstract}

\maketitle

\section{Introduction}

In principle, numerical simulations of simple liquids by the method of molecular dynamics (MD) is a quite old topic, for example, quite accurate estimations of details of molecular motion and the self-diffusion coefficient in comparison with the actual experimental data for liquid argon are dated back to the seminal work by A.~Rahman \cite{Rahman1964}; a review of the further developing of such simulations and their results can be found in Ref.~\cite{Meier2004}. 
However, their majority is limited by the vicinity of the liquid-vapor coexistence curve and moderately high supercritical pressures at low temperatures as well as time intervals corresponding to the stable normal diffusive regime. This fact may be conditioned by a limited number of experimental data available for simulations testing and discussion, especially liquified noble gases \cite{Barton1970,Suarez2015} as well as techniques for their obtaining. Thus, recent model studies utilizing MD simulations are shifted either to the region of supercritical and supercooled fluid states, where specific structure conditions lead to a variety of anomalous effects reflected in the transport coefficients even for relatively large characteristic times \cite{Baidakov2010,Ediger2012,Costigliola2016,Ohtori2017} or to a case of mixtures \cite{Thirumalai1989,Kob1995}, where different local spatial scales of interacting particles lead to similar effects. 

At the same time, the Lennard-Jones simple liquid can play the role of a model system even for discussing anomalous diffusion in more complex media mimicking the problems, which arise in biophysical systems \cite{Jeon2016,Ghosh2016}. In this case, extremely short-time range dynamics may be crucial since it strongly depends on the microscopic surrounding of a moving particle leading to different effects, such as particle population splitting, non-ergodicity, etc. \cite{Cherstvy2013,Schulz2013,Schulz2014}. Due to the existence of inter-particle interactions, such walking processes and structural features should be taken into account not only in the context of trapping but also reaction binding, see e.g. \cite{Grebenkov2018} including such a hot topic as the direct MD simulation of forming mesoscopic objects like viruses \cite{Tarasova2018}.
It also should be pointed out that dynamical features of particles motion in liquids are directly connected with their local microscopic structure of liquids, which determines rather complex transition from short- to long-times scales, as is has been revealed by considering a model system of hard spheres in \cite{Hopkins2010}.

But in contrast to the studies mentioned above, which were primarily addressed to simplified model systems, this work is intended to consider an interplay of structural, fluctuational and diffusional properties of simple liquid via a case study of liquid argon in the range of a condition resembling the experimentally accessible states as close as possible to the latter.  It is focused on the relatively unexplored in details of the short-time range of processes in this pure liquid under high pressure and low temperatures, where an influence of liquid's microscopic structure is sufficient. 

Respectively, changes in microstructure should be unavoidably reflected in as thermodynamic quantity such as the excess entropy, which determines the behavior of the macroscopic coefficient of self-diffusion \cite{Dyre2018}. Thus, the second part of this work deals with the large-time counterpart of the problem: an analytic predictive calculation of the self-diffusion coefficient in liquid argon for the same pressure-density-temperature conditions basing on an interplay between the density fluctuations and measurable thermodynamic quantities, the density and the isothermal compressibility. To assure physical relevance, the actual experimental information on thermodynamic and transport properties known in reference literature and databases are used for the direct comparison of calculated and measured data.

\section{Reduced density fluctuations and self-diffusion}

\subsection{Reduced fluctuations and self-diffusion along an isotherm and the saturation curve}

To analyze the pressure- and density-dependent behavior of the self-diffusion in the considered LJ-fluid simulating liquid argon, it is worthy to address the relative density fluctuations (a ratio of the actual density fluctuations to the density fluctuations in a hypothetical medium with the same values of thermodynamic parameters but in the state of the ideal gas):
\begin{equation}
\nu^{-1}=\left.\frac{\langle \left(\Delta \rho\right)^2 \rangle}{\rho}\right/
\left[\frac{\langle \left(\Delta \rho\right)^2 \rangle}{\rho}\right]_{i.g.}
=\frac{\mu_0}{RT}\rho \kappa_T,
\label{nu}
\end{equation}
which allows for  analysis of an interplay between microscopic liquid structure and thermodynamics \cite{Goncharov2013} since  $\kappa_T$,
$\mu_0$, $R$ are the isothermal compressibility,  the molar mass, and the gas constant, respectively. 

\begin{figure}%
\includegraphics[width=\columnwidth]{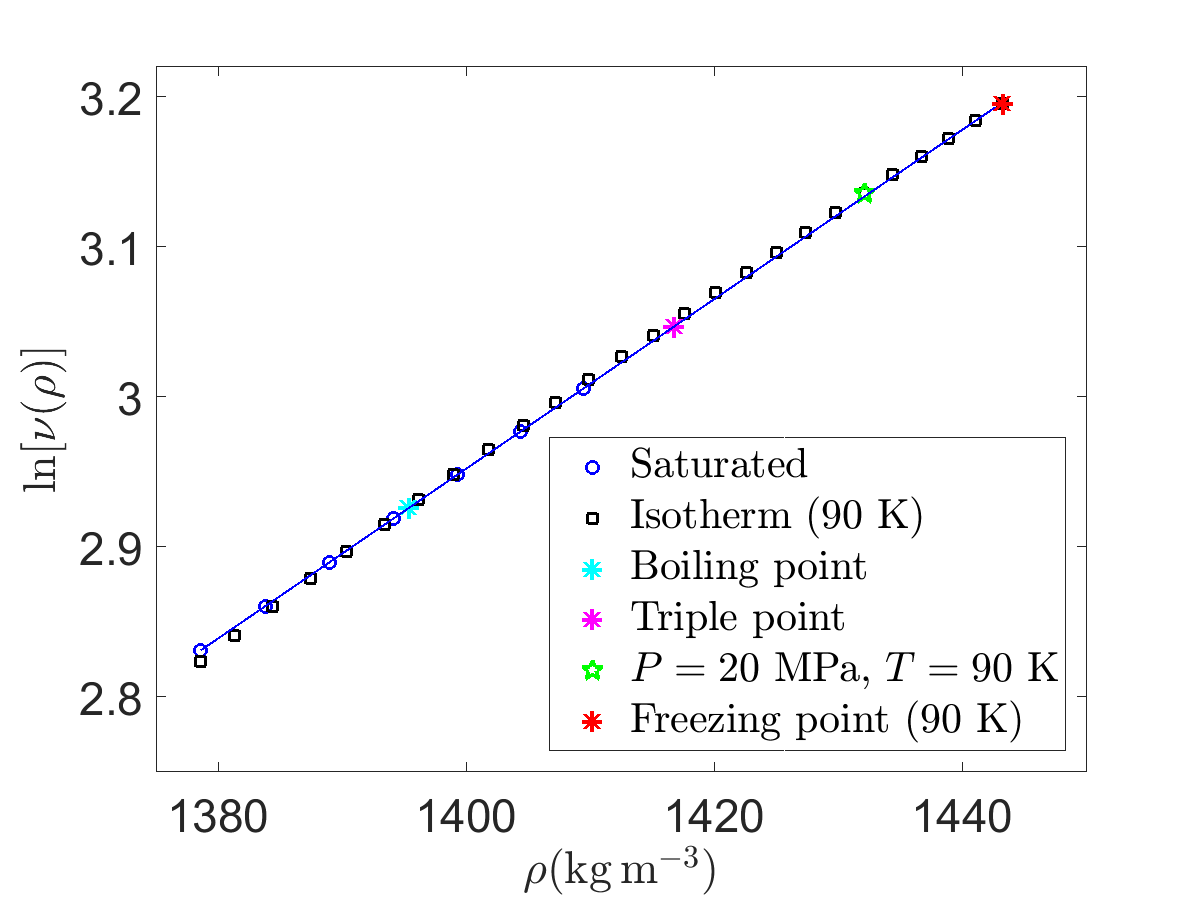}%
\caption{(Color online) The logarithm of inverse reduced density fluctuations (\ref{nu}) density calculated using actual thermodynamic  data of liquid argon at the saturation conditions from $T=90~\mathrm{K}$ to the triple point and along the isotherm $T=90~\mathrm{K}$. Thin solid line is a linear fit of the saturated data.
}%
\label{figFluct}%
\end{figure}

It is known that the logarithm of this parameter has a practically linear dependence on the density \cite{Goncharov2013,Chorazewski2017}, $\mathrm{ln}(\nu)=k\rho+b$, where $k$ and $b$ are substance-dependent parameters, the same for both the liquid-vapor coexisting curve and in the single-phase region (in the latter case this universality is violated under pressures higher than several hundred MPa only).
Figure~\ref{figFluct} illustrates this fact showing the logarithm of the reduced density fluctuations (\ref{nu}) as a function of the density simultaneously calculated along the saturation curve up to the triple point and along the isotherm $T=90~\mathrm{K}$ up to the vicinity of the freezing pressure at this temperature. The experiment-based database \cite{NIST} was used as a source of thermodynamic data, i.e. they are completely independent of simulations. 

At first, one can see that all the dots are placed along one practically straight line. 
At second, the sequence of blue circles corresponding to the simultaneous change of the density and the temperature along the saturation curve and the black squares corresponding to the pressure change along the isotherm overlap up to the density corresponding to the triple point, i.e. to the normal freezing density. But further, the set of squares continues this sequence along the same straight line calculated using the saturated data. Thus, from the point of view of such continuation, the isothermal fluctuations at high pressures should correspond to the fluctuations in metastable liquid argon under the normal saturated vapor, when argon it is accurately kept in a fluid non-frozen state, i.e. it represents a supercooled liquid.

Such behavior makes it possible to derive the so-called Fluctuation Theory-based Equation of State (FT-EoS)
\begin{equation}
\rho=\rho_0+\frac{1}{k}\mathrm{ln}\left[\frac{k\mu_0}{\nu RT}(P-P_0)+1\right],
\label{FT-EoS}
\end{equation}
which is based on the mentioned universality that makes available predicting the density of liquids, from simple to quite complex substances, see \cite{Chorazewski2017} and references therein, using the data measured at normal conditions only. This possibility originates from the more physically relevant picture of elastic properties of liquids in comparison, say, with the empiric Tait equation widely used for a pure fitting approach, as discussed in \cite{Chorazewski2017}.  

Figure~\ref{figRhoSD}(A) shows the predicted density of argon along the isotherm $T=90~\mathrm{K}$ from the liquid-vapor coexistence curve to the vicinity of the freezing point ($25~\mathrm{MPa}$) at this temperature; the average absolute deviation between them is equal to 0.0065~\%. Experimental data for the density under pressure as well as the saturated density, speed of sound and the heat capacity ratio used for computing the isothermal compressibility and, subsequently, $\nu$ and $k$ (see the description of the algorithm in Ref.~\cite{Chorazewski2017}) are taken from the database of the National Institute of Standards and Technology (NIST) \cite{NIST}, which is based on the high-accurate equation of state fitting a wide compilation of different experimental data \cite{Tegeler1999}. The numerical values of parameters used to obtain plots and the fitting procedure are given in the Appendix. 

\subsection{Molecular dynamics simulations}

Since density fluctuations in continuous media should be connected with the process of self-diffusion, consider some simulations related to the points of thermodynamics state discussed above.

The free available {\sc Python} code \cite{gitArgon} was used for simulations. It realizes the molecular dynamics algorithm
considering Newton's law dynamics
\begin{align}
\dot{\mathbf{r}_j}&=\mathbf{\mathbf{v}_j},\label{dynr}\\
\dot{\mathbf{v}_j}&=\frac{1}{m_0}\sum_{i\neq j}\mathbf{\mathbf{f}_{ij}},\label{dynv}
\end{align}
for $60912$ particles interacting by forces $\mathbf{f}=-\nabla U$ determined via the Lennard-Jones pair-wise potential 
$$
U(r)=4\epsilon\left[\left(\frac{\sigma}{r}\right)^{12}-\left(\frac{\sigma}{r}\right)^6\right],
$$
with the parameters $\epsilon/k_B=120~\mathrm{K}$, $\sigma=3.4~\mathrm{\AA}$ ($k_B$ is Boltzmann's constant) that corresponds to the classic Rahman's system \cite{Rahman1964} but extended in size ($864\times8$). Periodic boundary conditions were applied to the box of the side $L_{box}$, which was fixed and chosen in such a way that the mean density of particles within the box corresponded to the actual density of liquid argon at 90~K, i.e.
 $L_{box}=10.229\sigma$ respectively to the density $\rho_{sat}=1378.6~\mathrm{kg\, m^{-3}}$ on the saturation curve, and $L_{box}=10.0607\sigma$ respectively to the density $\rho_{20}=1432.1~\mathrm{kg\, m^{-3}}$ at 20~MPa on the isotherm. 
Note also that due to a limited number of experimental data on the self-diffusion coefficient of liquid argon, especially, under pressure for temperatures from a ``normal range'', i.e. sufficiently below the critical point \cite{Barton1970,Suarez2015}, the isotherm 90~K provides a most data-rich choice in a vicinity of the boiling point. In addition, the simulations were evaluated for the same density $\rho_{20}$ but the temperature $T=81.27~\mathrm{K}$, which was obtained as corresponding to this density in the supercooled (liquid) state along the saturation curve derived by extrapolating $\rho(T)$ dependence along this curve.

The potential's cut-off was chosen as 
$r_{cut} = 4.5\sigma$ that is twice of the conventional choice $r_{cut} = 2.25\sigma$ to assure possible long-ranged interaction in a high-density compressed liquid: the chosen value corresponds to 8 van der Waals radii $r_{W}=1.88~\mathrm{\AA}$ of the argon atom, i.e. to the fourth coordination sphere (although comparative studies showed that the results do not differ significantly, i.e. nearest-neighbor interactions between atoms prevails, as expected). The time step of simulations was equal to $dt=10^{-14}~\mathrm{s}$.

Before measurements, the system was equilibrating up to 470~ps; the equilibration was controlled by the plots of the pressure and the temperature of the system. The co-ordinates of all atoms were recorded for the subsequent 4096 time iterations (40.96~ps). To exclude effects of periodization, all trajectories were corrected via the shift of co-ordinates on $L_{box}$ when a particle reached the box's side and the respective co-ordinate jumps occurred. 

The time-averaged mean square displacement (tMSD) for each trajectory was calculated for $m$ elementary time steps $dt$ as
\begin{align}
\overline{\delta f^2}(\tau=mdt)=&\frac{1}{N-m}\sum_{i=1}^{N-m}\left[\left(f_x(t_i+mdt)-f_x(t_i)\right)^2+\right. \nonumber\\
&\left(f_y(t_i+mdt)-f_y(t_i)\right)^2+ \nonumber\\
&\left.\left(f_z(t_i+mdt)-f_z(t_i)\right)^2
 \right],
\label{defMSD}
\end{align}
where $f_{x,y,z}$ are components either of the displacement $\mathbf{r}$ (for the true MSD) or the velocity 
  $\mathbf{v}=(\mathbf{r}(t_i+mdt)-\mathbf{r}(t_i))(mdt)^{-1}$ (in this case, the mean square velocity, MSV, is considered).

Note that  all trajectories, which crossed the boundary and were continued under periodicity condition, were shifted and ``glued'' in these points of jumps in such a way that they went out on the initial box keeping the continuity of trajectories and velocities. The ensemble-averaged eMSD and eMSV were calculated for all cases indicating that they coincide with tMSD (the walks are ergodic), thus further the notations MSD and MSV will be used.

\begin{figure}%
\includegraphics[width=\columnwidth]{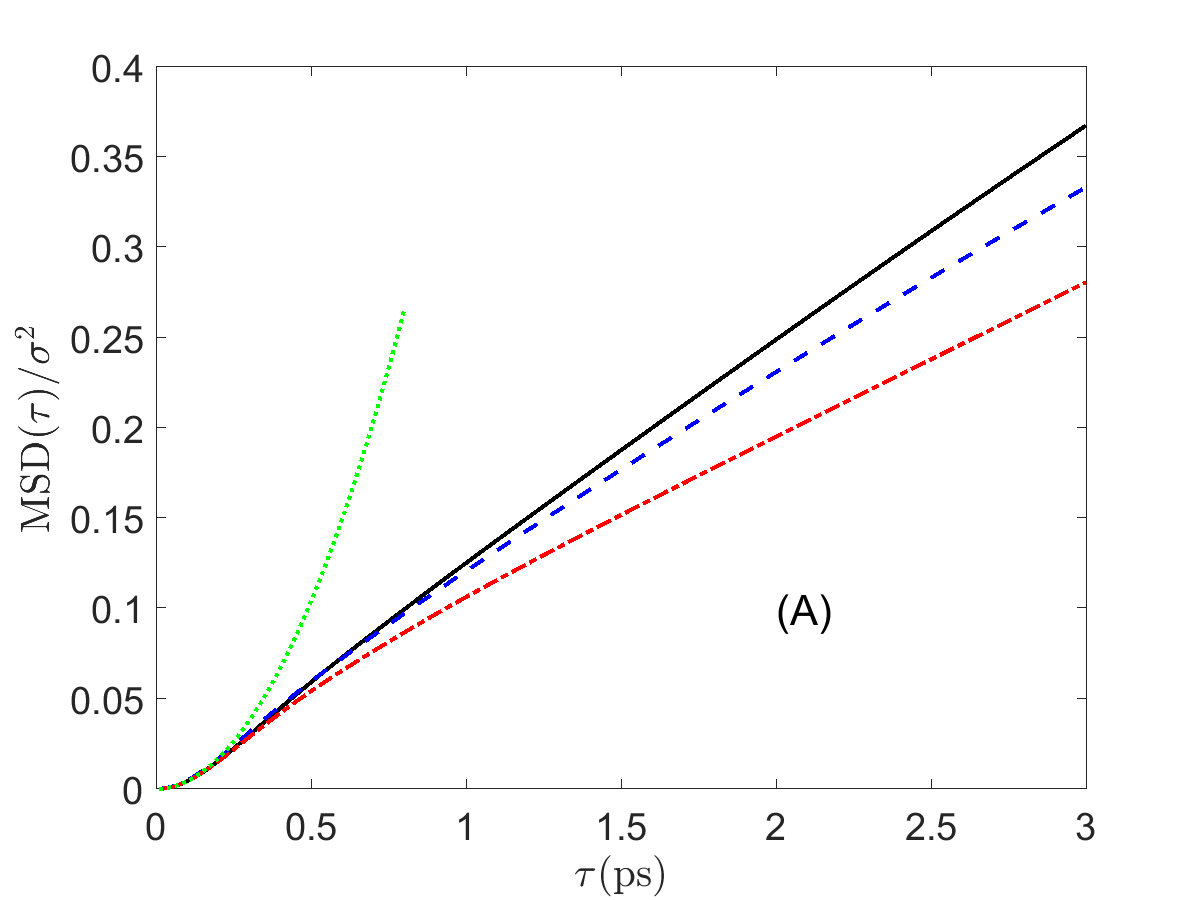}\\%
\includegraphics[width=\columnwidth]{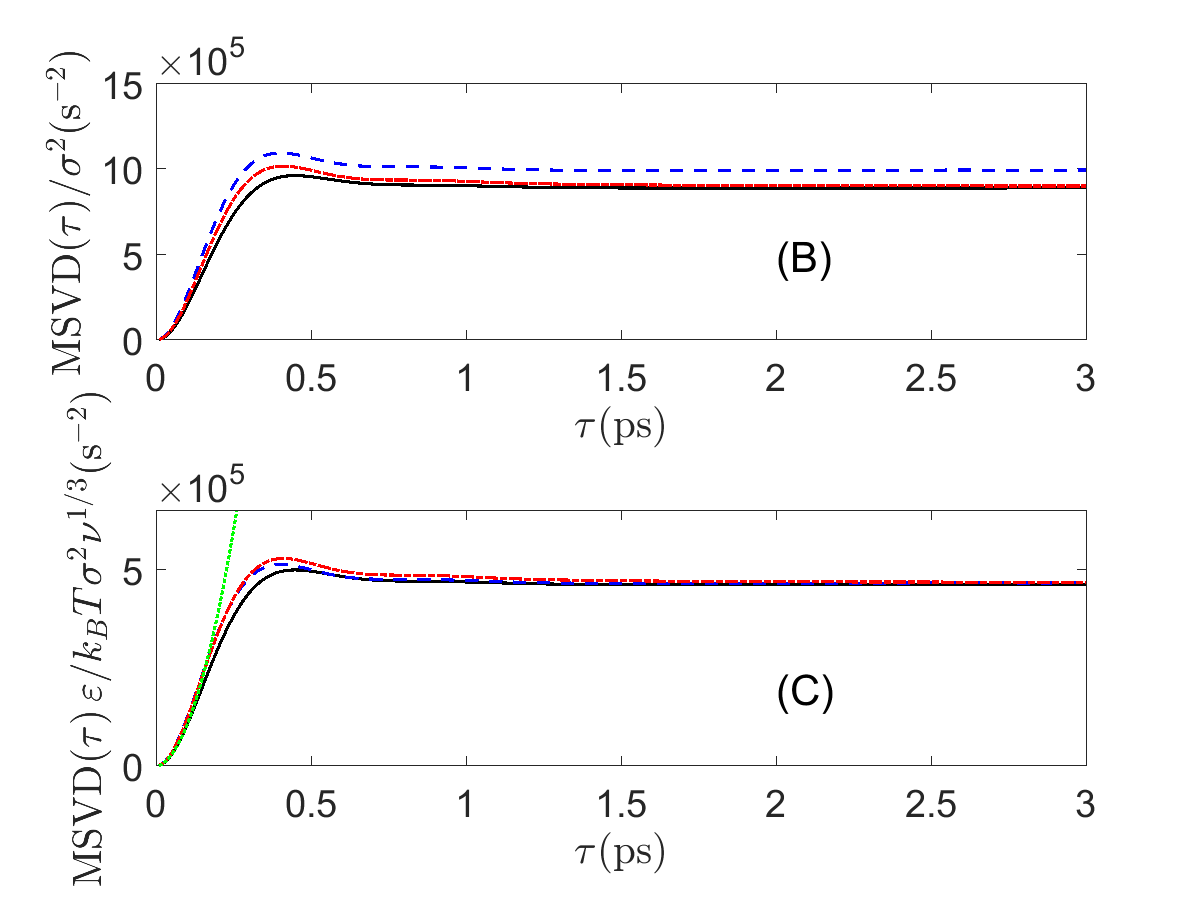}%
\caption{(Color online) Mean square displacements for the coordinates, i.e. Eq.~(\ref{defMSD}) with $f_{x,y,z}=(x,y,z)$ (MSD)  (A) and the velocity, i.e. Eq.~(\ref{defMSD}) with $f_{x,y,z}=(v_x,v_y,v_z)$ (MSV) with two kinds of scaling (B), (C)  in L-J liquids mimicking liquid argon at $T=90~\mathrm{K}$ under saturation conditions (black solid curve), $P=20~M\mathrm{MPa}$ (blue dashed curve), and the density corresponding to the latter at at $T=81.27~\mathrm{K}$ (red dash-dotted curve). The green dotted line fits a ballistic motion at very short times.}%
\label{figMSD}%
\end{figure}

Figure~\ref{figMSD}(A) shows the MSD for three considered conditions, where the length variable is rescaled to the characteristic length included into the L-J potential; note also that the value $2^{1/6}\sigma\approx 2r_W$, i.e. to the characteristic inter-particle distance ($r_W$ is the van der Waals radius of argon). The time variable is kept dimensional for a comparison with times, which are accessible in real physical experiments for studying the self-diffusion. 
 One can see that for short times the particle's motion is ballistic that is highlighted by the green dashed parabola. The respective displacements are quite short, of order $0.15~\sigma$ that correspond to motions just in a small vicinity of the potential's minimum, where forces acting on the particle is practically negligible. During the next time interval of around 1-2~ps, the MSD exhibits more sophisticated behavior corresponding to the crossover from quadratic to linear time dependence. The latter function corresponds to the normal diffusion and is detected for $\tau>2~\mathrm{ps}$. Different slopes for three different thermodynamic states indicate different values of the self-diffusion coefficient. To check the consistency of simulations, note that the diffusive regime for the saturated argon at $T=90~\mathrm{K}$ is found as
 $D_{sat}=2.3\, 10^{-9}~\mathrm{m^2\, t^{-1}}$ that corresponds to the value obtained by other authors for the L-J system mimicking liquid argon at the same temperature \cite{Wei2008} and belongs to the middle of the interval of its experimental values determined in different sources: 
$D_{sat}^{(exp)}=(1.89\pm 0.08)\, 10^{-9}~\mathrm{m^2\, t^{-1}}$, $D_{sat}^{(exp)}=(2.10\pm 0.10)\, 10^{-9}~\mathrm{m^2\, t^{-1}}$ \cite{Cini1960}, $D_{sat}^{(exp)}=2.43\, 10^{-9}~\mathrm{m^2\, t^{-1}}$ \cite{Naghizadeh1962}. More compressed liquid at the same temperature shows smaller self-diffusion, and the least one is the self-diffusion at the least temperature but the same density as the previous one. This is quite expectable, but one can draw more interesting conclusions about details of molecular motions using MSVD shown in Fig.~\ref{figMSD}(B),(C). 

First of all, Figure~\ref{figMSD}(B) indicates that the crossover from the ballistic to the diffusive motion in Figure~\ref{figMSD}(A) corresponds to the sufficiently non-monotonic behavior of MSVD in Figure~\ref{figMSD}(B). The boundary regimes are a ballistic parabola and a constant line, and there is a maximum in between. Physically it may be interpreted as a fast collision-reflection-permeation process when walking particles interact with particles, which comprise the first co-ordination shell. The value of this maximum is largest for the overdamped liquid at $T=90~\mathrm{K}$, i.e. particles move fast but they are sufficiently squeezed. The same squeezing for $T=81.27~\mathrm{K}$ results in a smaller ``hump'', and the smallest one corresponds to the saturated case of fast-moving particles, for which a larger free volume is available. All cases result in the constant MSVD, i.e. in the thermalization of motion, when the dispersion of velocities is fixed and the random walk is Brownian. 

All three MSVD curves are distinct after leaving the regime of the practically force-less ballistic flights in the vicinity of the potential well minimum. However, it is possible to rescale them to reveal a kind of universality, which reflects the basic physical premises of random displacements. First of all, magnitudes of the velocity should be determined by the system temperature as follows from the basics of statistical thermodynamics. Secondly, a particle gets a different velocity while it moves within an available free volume before the thermalization resulting in the diffusional motion. The respective statistics can be characterized by the reduced density fluctuation parameter (\ref{nu}), which is also equal to the reduced volume fluctuations. Since MSDV relates to the one-dimensional length units squared, the resulting dimensionless scaling factor takes the form $\varepsilon/k_BT\nu^{1/3}$, where $\varepsilon/k_B$ defines the characteristic L-J temperature. 

Figure~\ref{figMSD}(C) shows that such scaling already results in merging TMSD curves and this picture is different from the one depicted in Figure~\ref{figMSD}(B). Especially demonstrative is the quite accurate coincidence of blue dashed and red dash-dotted curves, which correspond to velocities in the supercooled and overdamped liquids that confirms fluctuation theory-based conclusions discussed above in relation to Figure~\ref{figFluct}. Both the ballistic (highlighted by the green dotted parabola) and the diffusional regimes are characterized by the same universal curve for all three cases. 

Such behavior of MSD and MSVD visible in Figure~\ref{figMSD} for an individual trajectory can be discussed in terms of random walks when the deterministic process defined by Newton's equations of motion (\ref{dynr})--(\ref{dynv}) is replaced with the system 
\begin{align}
\dot{\mathbf{r}}_j&=\mathbf{\mathbf{v}_j},\label{stochr}\\
\dot{\mathbf{v}}_j&=\frac{1}{m_0}\xi(t),\label{stochv}
\end{align}
where $\xi(t)$ is an appropriate random process. 

Note that such representation differs from a conventional discussion of random walks with the transition from ballistic to diffusional MSD introducing the underdamped Langevin equation with a Gaussian (white or colored) random noise and velocity-dependent damping \cite{Bodrova2016,Fa2018book}. The mentioned approach postulates a kind of random force and introduces the viscous damping from some macroscopic manifestations adjusted to the chosen noise character, while Eqs.~(\ref{stochr})--(\ref{stochv}) play a role of a stochastic counterpart to dynamical Eqs.~(\ref{dynr})--(\ref{dynv}), where the stochasticity originates from non-uniform distribution of nearest-neighbor particles acting on the observed one. This consideration provides a more direct interpretation for the behavior of MSD and MSDV shown in Fig~\ref{figMSD}, where the noise term in Eqs.~(\ref{stochv}) can be directly characterized by the recorded ensemble of trajectories. The most informative characteristic of such random forces (random accelerations) is its power spectral density (PSD) $\langle|\mathbf{F}_j(\omega)|^2\rangle$ over the period of observations $t_{obs}$ averaged over the ensemble of trajectories, where 
\begin{equation}
\mathbf{F}_j(\omega)=\frac{1}{3t_{obs}}\int_0^{t_{obs}}\dot{\mathbf{v}}_je^{i\omega t}dt.
\label{PSD}
\end{equation}

\begin{figure}%
\includegraphics[width=\columnwidth]{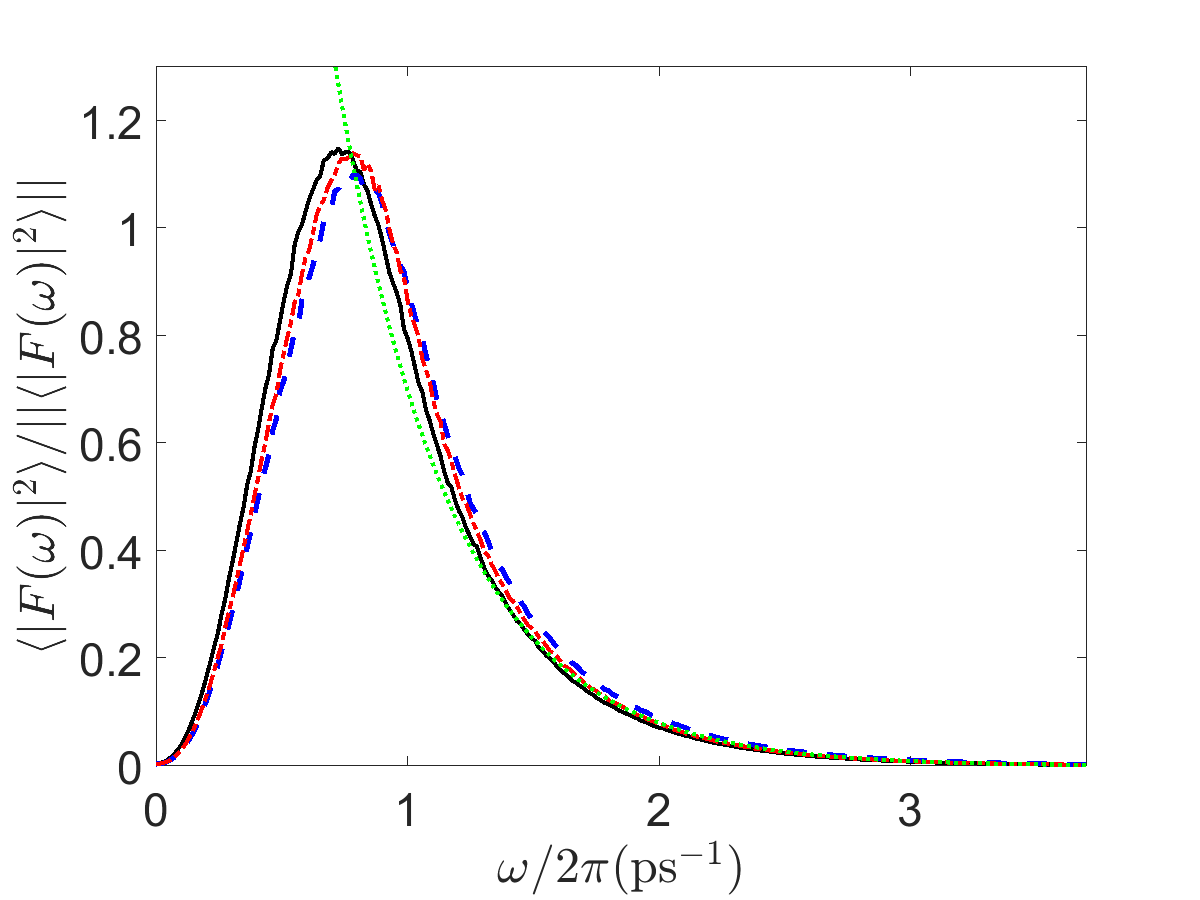}
\caption{(Color online) The normalized power spectral density of the random force $\xi(i)$ included into Eq.~(\ref{stochv}) for three cases considered -- their coloring is the same as in Figure~\ref{figMSD}. The green dotted line is an exponential function shown for guidance.}%
\label{figPSD}%
\end{figure}

Figure~\ref{figPSD} shows the averaged power spectral densities for three considered thermodynamic states (\ref{PSD}) normalized to the unit area under each curve. All of them have specific qualitative shape features, which explains the behavior of MSVD and MSD shown in Figure~\ref{figMSD}. 

The PSD curve has a clear maximum that indicates the existence of a leading oscillatory term; its period is equal to $2\pi/\omega_{max}=1.38~\mathrm{ps}$ for $\rho_{90}$ and $1.27~\mathrm{ps}$ for $\rho_{90}$, respectively. Note that these period values are coordinated approximately with the end of the transient process Figs.~\ref{figMSD}(B)--(C), i.e. larger time (and, respectively, spatial) scales are thermodynamic. In the  parts of the PSD for the frequencies $\omega/2\pi<0.5~\mathrm{ps}^{-1}$, i.e. to times larger that $2\,\mathrm{ps}$, blue dashed and red dash-dotted curves in Figure~\ref{figPSD} practically coincide (and they are quite close everywhere being distinct from the black solid line), that has correspondence to the equivalence of the slow density fluctuation processes under thermodynamic spatial and temporal scales as it is reflected in Figure~\ref{figFluct}. 
Finally, the PSD clearly tends to zero as $\omega\to 0$. Thus, the most probable particle's motion has an isotropic oscillatory character within cages of radii corresponding to average interparticle distance but existing density fluctuations leads to a wide distribution with respect to frequencies instead of a localized peak. 

This distribution, however, has a fast (exponential) decay (see the green guiding line in Figure~\ref{PSD}). Practically, there are no frequency components for $\omega/2\pi>3.5~\mathrm{ps}^{-1}$, i.e. oscillations with periods less than $\tau_{bound}=0.28~\mathrm{ps}$. Looking at Figs.~\ref{figMSD}(B)--(C), one can see that it is accurately the right boundary of ballistic behavior for the mean square velocity displacements, i.e. a practical absence of oscillatory components with higher frequencies results in the principal absence of disturbances leading to deviations from the linear dependence of MSDV on time for the walking process with zero mean
\begin{equation}
\mathbf{v}_j(t)=\frac{1}{m_0}\int_0^t\xi(t)dt=\frac{t_{obs}}{2\pi}\int_{-\infty}^{+\infty}\mathbf{F}_j(\omega)e^{-i\omega t}d\omega,
\label{smoothv}
\end{equation}

If one replaces the limits of the integral above by the values defined by $\tau_{bound}$, such velocity displacements will resemble so-called ``smooth random process'' extensively discussed recently in \cite{Filip2019} and in references therein. Such interpretation is well-fitted with the ideology of molecular dynamics since Eqs.~(\ref{dynr})--(\ref{dynv}) are dynamic, i.e. resulting trajectories are differentiable and their irregularity is based on the irregularity of surrounding particles acting on the observed particle. This is actually the case of Eqs.~(\ref{stochv}),~(\ref{smoothv}). In turn, the integration of Eq.~(\ref{smoothv}) as follows from Eq.~(\ref{stochr}) results in an irregular path. Taking into account times larger than $\tau_{bond}$, the velocities $\mathbf{v}_j(t)$ can be considered as effective random variables and their integration (a sequential summation) leads to the Brownian motion due to the Central Limit Theorem (there are no power-law tails here) with the MSD seen in Figure~\ref{figMSD}(A).

\subsection{Self-diffusion under high pressures via the thermodynamic fluctuations route}

Now it is possible to analyze the possibility of extending this fluctuation-based method from volumetric to transport properties, in particular, to self-diffusion in the thermodynamic limit of scales whose boundaries are revealed in the results of MD simulations above.  
The dimensionless reduced coefficient of self-diffusion in liquids in a majority of cases exhibit a universal exponential behavior \cite{Rosenfeld1977,Dzugutov1996,Rosenfeld1999,Dyre2018}
\begin{equation}
\left(\rho^{1/3}\sqrt{\mu_0/RT)}\right)D\propto e^{-A\frac{S_{ex}}{Nk_B}},
\label{defD}
\end{equation}
where $S_{ex}/Nk_B$ is the excess entropy per particle, which defines the difference between the entropy of the system under study and the entropy of an equivalent ideal gas at the same temperature and density, i.e. refers to the same pair of states as Eq.~(\ref{nu}); $A$ is a positive substance-dependent constant.  

It should be pointed out that the excess entropy is not a quantity, which can be determined straightforwardly in an experiment and its value depends on fitting to a chosen model \cite{Dyre2018}. In particular, an approach relatively easily applicable to results of numerical simulations replaces the full excess entropy by the pair entropy \cite{Baranyai1989,Saija1996},  which is, in fact, the leading term in the expansion of the complete thermodynamic function into the Taylor series with respect to contributions of particles pairs, triplets, etc.

But the simplest possibility, valid not only in the low-density limit, is a usage of  purely thermodynamic quantities, namely the compressibility factor 
$Z=\mu_0P/\rho RT$ and its isochoric derivative \cite{Vaz2012}
\begin{equation}
\frac{S_{ex}}{Nk_B}=-\int\limits_0^{\rho}\left[T\left(\frac{\partial Z}{\partial T}\right)_V+Z(\rho)-1\right]\frac{d\rho}{\rho}.
\label{Stherm}
\end{equation}

\begin{figure}%
\includegraphics[width=\columnwidth]{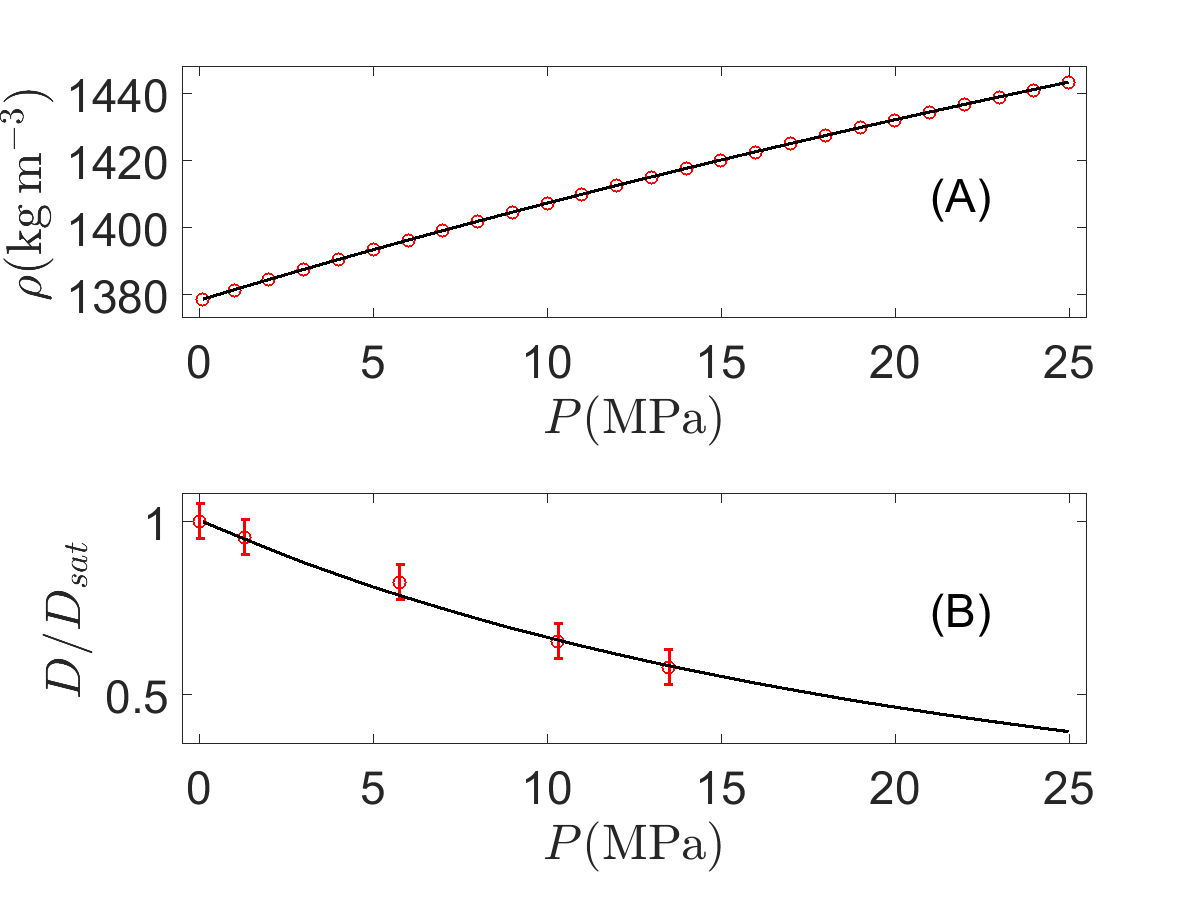}%
\caption{(Color online) Experimental (red circles) and calculated via FT-EoS values of the density (A) and the reduced self-diffusion coefficient (B) of liquid argon along the isotherm $T=90~\mathrm{K}$. For the density, the experimental data uncertainty range does not exceed markers size; for the coefficient of self-diffusion, it is denoted explicitly. 
}%
\label{figRhoSD}%
\end{figure}

Using the standard thermodynamic definition of the internal pressure 
$$
P_i=T\left(\frac{\partial P }{\partial T}\right)_V-P,
$$
the derivative mentioned above can be easily calculated and Eq.~(\ref{Stherm}) rewritten as 
\begin{equation}
\frac{\Delta S_{ex}'}{k_B}=-\int\limits_{\rho_{0}}^{\rho}\left[\frac{\mu_0 P_i}{\rho RT}+\frac{\mu_0P(\rho)}{\rho RT}-1\right]\frac{d\rho}{\rho}
\label{SthermP}
\end{equation}
if one considers a difference between excess entropies of two states with the densities $\rho_0$ and $\rho$ placed on one isotherm. After substituting into the integrand the pressure expressed from FT-EoS (\ref{FT-EoS})  as an explicit function of the density, the integral (\ref{SthermP}) is taken analytically that results in
\begin{align}
\frac{\Delta S_{ex}'}{k_B}=&-\left[
-\frac{\mu_0 (P_i+P_0)}{ RT \rho}+\frac{\nu}{k\rho}-\mathrm{ln}(k\rho)\right. \nonumber\\
&+
\left.\left.\nu e^{-k\rho_s}\left(\mathrm{Ei}(k\rho)-\frac{e^{k\rho}}{k\rho}\right)
\right]\right|_{\rho_{0}}^{\rho},
\label{aSanalyt}
\end{align}
where $\mathrm{Ei}(k\rho)$ is the exponential integral, and $P_i=\mathrm{const}$ as it follows from the basic construction of FT-EoS (\ref{FT-EoS}); although it fulfils only approximately in average, the actual changes of $P_i$ with respect to its value within the considered range of pressures can be neglected. Accuracy of the density prediction shown in Figure~\ref{figRhoSD}(A) confirms this. 

Figure~\ref{figRhoSD}(B) shows the curve of the relative coefficient of self-diffusion change calculated using
 Eqs.~(\ref{aSanalyt}),~(\ref{defD}) with $A=3.2$ in comparison with the raw experimental ratio of the coefficient of self-diffusion under pressure to its saturated value taken from \cite{Naghizadeh1962}.  One can see that the proposed density fluctuation-based model quite accurately reproduces the non-linear character of the self-diffusion coefficient's diminishing with the growing pressure.

\subsection{Particle number distribution as an exponential dispersion model}

Now let us consider this model from the point of view of the probability density functions (p.d.f.) for particles located within a small (but not too small, see below) volume surrounding a chosen particle. This analysis can be done following the approach called an exponential dispersion model (EDM) \cite{Jorgensen1997book} describing statistical distributions for which the variance of a random variable is a function of its mean value. 

Such property is fulfilled, when the p.d.f. has a form 
\begin{equation}
w(N,\theta)=a(N)\exp\left[N\theta-\kappa(\theta)\right].
\label{w}
\end{equation}
Here it is written with respect to the number of particles $N$; $\kappa(\theta)$ is the cumulant function  defined in a standard way as the natural logarithm of the moment-generating function (moments are denoted as $M_j$ here)
$$
\kappa(\theta)=\mathrm{ln}\mathbf{E}\left(e^{\theta N}\right)=\sum\limits_{j=1}^{\infty}M_j\frac{\theta^j}{j!},
$$
i.e. coefficients of its expansion into the Taylor series are the moments of distribution; the parameter $\theta$, with respect to which this expansion is written, is called the the canonical parameter; and $a(N)$ is some suitable function, which assures the norm of this distribution. 

Following this definition, the mean number of particles  within the chosen volume and its variance are determined as
\begin{equation}
\bar{N}=\frac{d\kappa}{d\theta}
\label{eq:mean}
\end{equation}
and 
\begin{equation}
\mathrm{var}(N)=\overline{(\Delta N)^2}=\frac{d^2\kappa}{d\theta^2}=\frac{d\bar{N}}{d\theta}.
\label{eq:var}
\end{equation}

Consider the volume as small but sufficiently macroscopic, i.e. one can express the number fluctuations in a fixed volume via the standard statistical thermodynamics expression \cite{Landau2013statphys}
\begin{equation}
\frac{\langle (\Delta N)^2\rangle}{N}=\frac{N}{V}k_BT\kappa_T,
\label{flucN}
\end{equation}
where $k_B$ is Boltzmann's constant. Since we operate with the macroscopic volume, for which one can determine the isothermal compressibility, the medium can be considered as uniform at such scales, i.e. $N/V\cong \langle N \rangle/V\equiv \bar{n}=\rho/m$, where $m$ is the mass of one particle. From the microscopical point of view, this condition is fulfilled when the volume's radius exceeds 3-4 mean inter-particle distances, when the radial distribution function (r.d.f.) approaches a horizontal line and, respectively, the number of particles inside a selected volume will grow with the growth of the latter linearly as proportional to the thermodynamic density. 

Under this assumption, Eq.~(\ref{flucN}) takes the form
$$
\frac{\langle (\Delta N^2)\rangle}{N}=\frac{RT}{\mu_0}\rho \kappa_T
$$
coinciding with Eq.~(\ref{nu}). For ideal gas the right-hand side is equal to 1, otherwise it has a form of the exponential function
$\exp\left[-(k\rho+b)\right]=\exp\left[-(km\bar{n}+b)\right]$. Respectively, multiplying the nominator and denominator in the left-hand side by the fixed volume squared $V^2$ and applying the same assumption that actual number density is equal to the mean number density, it is possible to conclude that the system satisfies the conditions of the EDM:
\begin{equation}
\overline{(\Delta n)^2}=(\nu_0V)^{-1}\bar{n}e^{-km\bar{n}},
\label{nun}
\end{equation}
where the notation $\nu_0=\exp(kb)$ is introduced. 

Therefore, the mean value satisfies the ordinary differential equation
$$
\frac{d\bar{n}}{d\theta}=(\nu_0V)^{-1}\bar{n}e^{-km\bar{n}},
$$
which can be easily solved by the variable separation method:
\begin{equation}
\theta=\nu_0V\mathrm{Ei}(km\bar{n})+c_1.
\label{partheta}
\end{equation}

Within the same way, it is possible to find the cumulant function using Eq.~(\ref{eq:mean}):
$$
\frac{d\kappa}{d\theta}=\frac{d\kappa}{d\bar{n}}\frac{d\bar{n}}{d\theta}=
\frac{d\kappa}{d\bar{n}}(\nu_0V)^{-1}\bar{n}e^{-km\bar{n}}=\bar{n},
$$
i.e.
$$
\frac{d\kappa}{d\bar{n}}=\nu_0\bar{n}e^{k\bar{n}}
$$
with the solution
\begin{equation}
\kappa(\bar{n})=\nu_0V(km)^{-1}e^{km\bar{n}}+c_2.
\label{parkappa}
\end{equation}

Note that $c_1$ and $c_2$ are additive constants, and a substitution of the expressions (\ref{partheta}), (\ref{parkappa})  into Eq.~(\ref{w}) will change the latter by a constant factor $\exp(c_2)$ and the still indefinite function $a(n)$ is the $n$-only dependent multiplier. Therefore, one can put both $c_1=0$ and $c_2=0$  without loss of generality. 

Thus, the conclusion is that the cumulant function in this case is the derivative of the reduced bulk modulus with respect to the density:
with the solution
$$
\kappa(\bar{n})=\nu_0V(km)^{-1}e^{km\bar{n}}=(km)^{-1}e^{km\bar{n}+b}=\frac{d\nu}{d\bar{n}}.
$$

The resulting p.d.f. has a form
\begin{equation}
w(n,\bar{n})=a(n)\exp\left[n\nu_0V\mathrm{Ei}(km\bar{n})-\nu_0V(km)^{-1}e^{km\bar{n}}\right].
\label{pdf}
\end{equation}
Note that it can be rewritten back to the number of particles $N=nV$ and the thermodynamic mass density $\rho=m\bar{n}=mn$, in the form independent on a particular value of the volume $V$ (keeping it larger then the region of r.d.f.'s oscillations)
$$
w(N,\rho)=a(N)\exp\left[N\left(\nu_0\mathrm{Ei}(k\rho)-\nu_0(k\rho)^{-1}e^{k\rho}\right)\right],
$$
i.e. the exponential term of the EDM's p.d.f. can be associated with a part in the thermodynamic entropic contribution into the self-diffusion (\ref{aSanalyt}), which contains the same combination of the exponential integral and the exponential function. 

However, it can be shown that Eq.~(\ref{pdf}) has a more wide applicability and can be used not only in the case of high (liquid) densities that can be demonstrated via a transition to the limiting case of extremely small number densities, i.e.,  $|k|\rho<<1$. This may be achieved considering 
the vapor branch at almost ideal gas conditions, when $\nu_0\approx 1$,  and $k<0$, see \cite{Goncharov2013}, where the inverse reduced density fluctuations under such conditions were studied based on actual experimental data for gaseous argon. 

In the considered case $\exp(k\rho)\approx 1+k\rho=1+km\bar{n}$, and
the cumulant function
$$
V(km)^{-1}e^{km\bar{n}}\approx V(km)^{-1}+\bar{N}
$$
since $\bar{n}=\bar{N}/V$.

The series representation of the exponential integral function in this case is 
$$
\mathrm{Ei}(km\bar{n})=\mathrm{ln}(\bar{N})+\left[\mathrm{ln}(km/V)-i\pi\right]+\gamma+
\sum_{j=1}^{\infty}\frac{(km\bar{n})^j}{j!j},
$$
where $\gamma$ is Euler's constant, and $\left[\mathrm{ln}(km/V)-i\pi\right]=\mathrm{Re}\mathrm{ln}(km/V)$ is a real number.

Where, 
\begin{multline*}
N\mathrm{Ei}(km\bar{n})=N\mathrm{ln}(\bar{N})+N\left[\gamma n+\mathrm{Re}\left\{\mathrm{ln}(km/V)\right\}\right]+\\
n\sum_{j=1}^{\infty}\frac{(km\bar{n})^j}{j!j}.
\end{multline*}

The exponential of the first term is equal to
$$
\exp\left(N\mathrm{ln}(\bar{N})\right)=\bar{N}^N,
$$
the last term can be neglected as small, the second term does not depend on $\bar{N}$ and, therefore, can be combined with $a(N)$.  Thus, the resulting p.d.f. is
$$
w(N)=\left\{a(N)e^{(\gamma+\mathrm{Re}\mathrm{ln}(km/V))N-V(km)^{-1}} \right\}\bar{N}^Ne^{-\bar{N}}.
$$
Taking the first factor as $1/N!$ [it is possible since $a(N)$ is a still indefinite function and it is required to norm the p.d.f. to unity], this expression reduces to the Poisson distribution 
$$
w(N)=\frac{\bar{N}^N}{N!}e^{-\bar{N}}
$$
that is known for the ideal gas (or in the case of small volumes containing an extremely small number of particles inside) \cite{Smoluchowski1904,Landau2013statphys}. 

Thus, the revealed non-trivial dependence on the liquid density, which can be connected with average inter-particle distances, may serve for a future microscopic probabilistic background for the studying transport processes in liquids under pressure, which exhibit a non-trivial behavior.

\section{Conclusion}

The main results of this work can be summarized as follows. Although the self-diffusion in simple liquids in the thermodynamic limit behaves like a normal diffusion process, its detailed picture for short times is more complex and exhibits properties of a random mixture of oscillatory modes whose frequencies are defined by free volume available for particles caged within their first coordination shells. The properties of such oscillations can be highlighted in the most direct way by the exploration of the mean square velocity displacement and connected with the thermodynamic density fluctuations. Note also that such behavior can be associated with the recent topic of non-stationary transient modes in model Ornstein-Uhlenbeck and related stochastic processes \cite{Thiel2016,Cherstvy2018,Kusmierz2018}. In addition, the approach considered in this work is not limited by the particular L-J potential only, and may be applied to systems with other potentials, e.g. in colloidal systems \cite{Pellicane2012}. This is because it addresses the density (or local volume) fluctuations, and, respectively, the liquid's structure factor, which can be calculated as a statistical quantity for liquid with various kinds of interactions. 

Finally, it should be pointed out that some kind of statistical equivalence of liquids behavior under elevated pressures or lowered temperatures can be described in a uniform way within a frame of the generalized linear model. This follows from a response of the density fluctuations on the growing particles packing that lead to the same value of the reduced density fluctuations. As a result,  an analytic predictive expression for calculating the density and the self-diffusion coefficient with an accuracy corresponding to the experimental one is available. Thus, this open perspective for analytical prediction of this parameter not only for argon but for more complex substances, e.g. organic molecular liquids, since the basic reference data can be extracted from thermodynamic databases.

\section*{ Acknowledgment} 

This work is partially supported by the National
Science Centre (Poland) Grant No. 2016/23/B/ST8/02968.

\section*{Appendix}

The density and the natural logarithm of the isothermal compressibility are fitted by polynomials using the standard {\sc MATLAB}'s function {\sc polyfit} \cite{polyfit} as  
$$
\tilde{f}(T)=\sum_{j=0}^{3}\tilde{F}_j\left(\frac{T-\tilde{T}}{\sigma_T}\right)^j,
$$
where $\tilde{f}$ is either $\rho$ measured in $\mathrm{kg\, m^{-3}}$ or $\mathrm{ln}(\kappa_T)$ measured in $\mathrm{Pa}^{-1}$, the temperature is in Kelvins. The respective coefficients are given in Table.~\ref{partable}.

\begin{table}[h!]
\caption{Parameters of the fitted thermodynamic quantities}
\label{partable}
\begin{ruledtabular}
\begin{tabular}{lrrrrrr}
& $\tilde{F}_3$ & $\tilde{F}_2$ & $\tilde{F}_1$ & $\tilde{F}_0$ & $\tilde{T}$ & $\sigma_T$\\
\hline
$\rho$ & -0.05600	& -0.63333 &-34.387	& 1362.8 & 92.5 & 5.40\\
$\mathrm{ln}(\kappa_T)$ &$1.613\times 10^{-4}$ 	&$5.721\times 10^{-2}$  &0.1611	& -19.82 &92.5 & 5.40\\
\end{tabular}
\end{ruledtabular}
\end{table}

For the key dimensionless fluctuation parameter, Eq.~(\ref{nu}), its natural logarithm  was linearly fitted as $\mathrm{ln}(\nu)=k\rho+b$ that gave  values of the coefficients as $k=0.00564~\mathrm{m^3\, kg^{-1}}$ and $b=-4.94$.


\begin{thebibliography}{44}%
\makeatletter
\providecommand \@ifxundefined [1]{%
 \@ifx{#1\undefined}
}%
\providecommand \@ifnum [1]{%
 \ifnum #1\expandafter \@firstoftwo
 \else \expandafter \@secondoftwo
 \fi
}%
\providecommand \@ifx [1]{%
 \ifx #1\expandafter \@firstoftwo
 \else \expandafter \@secondoftwo
 \fi
}%
\providecommand \natexlab [1]{#1}%
\providecommand \enquote  [1]{``#1''}%
\providecommand \bibnamefont  [1]{#1}%
\providecommand \bibfnamefont [1]{#1}%
\providecommand \citenamefont [1]{#1}%
\providecommand \href@noop [0]{\@secondoftwo}%
\providecommand \href [0]{\begingroup \@sanitize@url \@href}%
\providecommand \@href[1]{\@@startlink{#1}\@@href}%
\providecommand \@@href[1]{\endgroup#1\@@endlink}%
\providecommand \@sanitize@url [0]{\catcode `\\12\catcode `\$12\catcode
  `\&12\catcode `\#12\catcode `\^12\catcode `\_12\catcode `\%12\relax}%
\providecommand \@@startlink[1]{}%
\providecommand \@@endlink[0]{}%
\providecommand \url  [0]{\begingroup\@sanitize@url \@url }%
\providecommand \@url [1]{\endgroup\@href {#1}{\urlprefix }}%
\providecommand \urlprefix  [0]{URL }%
\providecommand \Eprint [0]{\href }%
\providecommand \doibase [0]{http://dx.doi.org/}%
\providecommand \selectlanguage [0]{\@gobble}%
\providecommand \bibinfo  [0]{\@secondoftwo}%
\providecommand \bibfield  [0]{\@secondoftwo}%
\providecommand \translation [1]{[#1]}%
\providecommand \BibitemOpen [0]{}%
\providecommand \bibitemStop [0]{}%
\providecommand \bibitemNoStop [0]{.\EOS\space}%
\providecommand \EOS [0]{\spacefactor3000\relax}%
\providecommand \BibitemShut  [1]{\csname bibitem#1\endcsname}%
\let\auto@bib@innerbib\@empty
\bibitem [{\citenamefont {Rahman}(1964)}]{Rahman1964}%
  \BibitemOpen
  \bibfield  {author} {\bibinfo {author} {\bibfnamefont {A.}~\bibnamefont
  {Rahman}},\ }\href@noop {} {\bibfield  {journal} {\bibinfo  {journal}
  {Physical Review}\ }\textbf {\bibinfo {volume} {136}},\ \bibinfo {pages}
  {A405} (\bibinfo {year} {1964})}\BibitemShut {NoStop}%
\bibitem [{\citenamefont {Meier}\ \emph {et~al.}(2004)\citenamefont {Meier},
  \citenamefont {Laesecke},\ and\ \citenamefont {Kabelac}}]{Meier2004}%
  \BibitemOpen
  \bibfield  {author} {\bibinfo {author} {\bibfnamefont {K.}~\bibnamefont
  {Meier}}, \bibinfo {author} {\bibfnamefont {A.}~\bibnamefont {Laesecke}}, \
  and\ \bibinfo {author} {\bibfnamefont {S.}~\bibnamefont {Kabelac}},\
  }\href@noop {} {\bibfield  {journal} {\bibinfo  {journal} {Journal of
  Chemical Physics}\ }\textbf {\bibinfo {volume} {121}},\ \bibinfo {pages}
  {9526} (\bibinfo {year} {2004})}\BibitemShut {NoStop}%
\bibitem [{\citenamefont {Barton}\ and\ \citenamefont
  {Speedy}(1970)}]{Barton1970}%
  \BibitemOpen
  \bibfield  {author} {\bibinfo {author} {\bibfnamefont {A.~F.~M.}\
  \bibnamefont {Barton}}\ and\ \bibinfo {author} {\bibfnamefont {R.~J.}\
  \bibnamefont {Speedy}},\ }\href@noop {} {\bibfield  {journal} {\bibinfo
  {journal} {High Temperatures -- High Pressures}\ }\textbf {\bibinfo {volume}
  {2}},\ \bibinfo {pages} {117} (\bibinfo {year} {1970})}\BibitemShut {NoStop}%
\bibitem [{\citenamefont {Su\'arez-Iglesias}\ \emph {et~al.}(2015)\citenamefont
  {Su\'arez-Iglesias}, \citenamefont {Medina}, \citenamefont {de~los
  Angeles~Sanz}, \citenamefont {Pizarro},\ and\ \citenamefont
  {Bueno}}]{Suarez2015}%
  \BibitemOpen
  \bibfield  {author} {\bibinfo {author} {\bibfnamefont {O.}~\bibnamefont
  {Su\'arez-Iglesias}}, \bibinfo {author} {\bibfnamefont {I.}~\bibnamefont
  {Medina}}, \bibinfo {author} {\bibfnamefont {M.}~\bibnamefont {de~los
  Angeles~Sanz}}, \bibinfo {author} {\bibfnamefont {C.}~\bibnamefont
  {Pizarro}}, \ and\ \bibinfo {author} {\bibfnamefont {J.~L.}\ \bibnamefont
  {Bueno}},\ }\href@noop {} {\bibfield  {journal} {\bibinfo  {journal} {Journal
  of Chemical \& Engineering Data}\ }\textbf {\bibinfo {volume} {60}},\
  \bibinfo {pages} {2757} (\bibinfo {year} {2015})}\BibitemShut {NoStop}%
\bibitem [{\citenamefont {Baidakov}\ and\ \citenamefont
  {Kozlova}(2010)}]{Baidakov2010}%
  \BibitemOpen
  \bibfield  {author} {\bibinfo {author} {\bibfnamefont {V.~G.}\ \bibnamefont
  {Baidakov}}\ and\ \bibinfo {author} {\bibfnamefont {Z.~R.}\ \bibnamefont
  {Kozlova}},\ }\href@noop {} {\bibfield  {journal} {\bibinfo  {journal}
  {Chemical Physics Letters}\ }\textbf {\bibinfo {volume} {500}},\ \bibinfo
  {pages} {23} (\bibinfo {year} {2010})}\BibitemShut {NoStop}%
\bibitem [{\citenamefont {Ediger}\ and\ \citenamefont
  {Harrowell}(2012)}]{Ediger2012}%
  \BibitemOpen
  \bibfield  {author} {\bibinfo {author} {\bibfnamefont {M.~D.}\ \bibnamefont
  {Ediger}}\ and\ \bibinfo {author} {\bibfnamefont {P.}~\bibnamefont
  {Harrowell}},\ }\href@noop {} {\bibfield  {journal} {\bibinfo  {journal}
  {Journal of Chemical Physics}\ }\textbf {\bibinfo {volume} {137}},\ \bibinfo
  {pages} {080901} (\bibinfo {year} {2012})}\BibitemShut {NoStop}%
\bibitem [{\citenamefont {Costigliola}\ \emph {et~al.}(2016)\citenamefont
  {Costigliola}, \citenamefont {Schr{\o}der},\ and\ \citenamefont
  {Dyre}}]{Costigliola2016}%
  \BibitemOpen
  \bibfield  {author} {\bibinfo {author} {\bibfnamefont {L.}~\bibnamefont
  {Costigliola}}, \bibinfo {author} {\bibfnamefont {T.~B.}\ \bibnamefont
  {Schr{\o}der}}, \ and\ \bibinfo {author} {\bibfnamefont {J.~C.}\ \bibnamefont
  {Dyre}},\ }\href@noop {} {\bibfield  {journal} {\bibinfo  {journal} {Physical
  Chemistry Chemical Physics}\ }\textbf {\bibinfo {volume} {18}},\ \bibinfo
  {pages} {14678} (\bibinfo {year} {2016})}\BibitemShut {NoStop}%
\bibitem [{\citenamefont {Ohtori}\ \emph {et~al.}(2017)\citenamefont {Ohtori},
  \citenamefont {Miyamoto},\ and\ \citenamefont {Ishii}}]{Ohtori2017}%
  \BibitemOpen
  \bibfield  {author} {\bibinfo {author} {\bibfnamefont {N.}~\bibnamefont
  {Ohtori}}, \bibinfo {author} {\bibfnamefont {S.}~\bibnamefont {Miyamoto}}, \
  and\ \bibinfo {author} {\bibfnamefont {Y.}~\bibnamefont {Ishii}},\
  }\href@noop {} {\bibfield  {journal} {\bibinfo  {journal} {Physical Review
  E}\ }\textbf {\bibinfo {volume} {95}},\ \bibinfo {pages} {052122} (\bibinfo
  {year} {2017})}\BibitemShut {NoStop}%
\bibitem [{\citenamefont {Thirumalai}\ \emph {et~al.}(1989)\citenamefont
  {Thirumalai}, \citenamefont {Mountain},\ and\ \citenamefont
  {Kirkpatrick}}]{Thirumalai1989}%
  \BibitemOpen
  \bibfield  {author} {\bibinfo {author} {\bibfnamefont {D.}~\bibnamefont
  {Thirumalai}}, \bibinfo {author} {\bibfnamefont {R.~D.}\ \bibnamefont
  {Mountain}}, \ and\ \bibinfo {author} {\bibfnamefont {T.~R.}\ \bibnamefont
  {Kirkpatrick}},\ }\href@noop {} {\bibfield  {journal} {\bibinfo  {journal}
  {Physical Review A}\ }\textbf {\bibinfo {volume} {39}},\ \bibinfo {pages}
  {3563} (\bibinfo {year} {1989})}\BibitemShut {NoStop}%
\bibitem [{\citenamefont {Kob}\ and\ \citenamefont {Andersen}(1995)}]{Kob1995}%
  \BibitemOpen
  \bibfield  {author} {\bibinfo {author} {\bibfnamefont {W.}~\bibnamefont
  {Kob}}\ and\ \bibinfo {author} {\bibfnamefont {H.~C.}\ \bibnamefont
  {Andersen}},\ }\href@noop {} {\bibfield  {journal} {\bibinfo  {journal}
  {Physical Review E}\ }\textbf {\bibinfo {volume} {51}},\ \bibinfo {pages}
  {4626} (\bibinfo {year} {1995})}\BibitemShut {NoStop}%
\bibitem [{\citenamefont {Jeon}\ \emph {et~al.}(2016)\citenamefont {Jeon},
  \citenamefont {Javanainen}, \citenamefont {Martinez-Seara}, \citenamefont
  {Metzler},\ and\ \citenamefont {Vattulainen}}]{Jeon2016}%
  \BibitemOpen
  \bibfield  {author} {\bibinfo {author} {\bibfnamefont {J.-H.}\ \bibnamefont
  {Jeon}}, \bibinfo {author} {\bibfnamefont {M.}~\bibnamefont {Javanainen}},
  \bibinfo {author} {\bibfnamefont {H.}~\bibnamefont {Martinez-Seara}},
  \bibinfo {author} {\bibfnamefont {R.}~\bibnamefont {Metzler}}, \ and\
  \bibinfo {author} {\bibfnamefont {I.}~\bibnamefont {Vattulainen}},\
  }\href@noop {} {\bibfield  {journal} {\bibinfo  {journal} {Physical Review
  X}\ }\textbf {\bibinfo {volume} {6}},\ \bibinfo {pages} {021006} (\bibinfo
  {year} {2016})}\BibitemShut {NoStop}%
\bibitem [{\citenamefont {Ghosh}\ \emph {et~al.}(2016)\citenamefont {Ghosh},
  \citenamefont {Cherstvy}, \citenamefont {Grebenkov},\ and\ \citenamefont
  {Metzler}}]{Ghosh2016}%
  \BibitemOpen
  \bibfield  {author} {\bibinfo {author} {\bibfnamefont {S.~K.}\ \bibnamefont
  {Ghosh}}, \bibinfo {author} {\bibfnamefont {A.~G.}\ \bibnamefont {Cherstvy}},
  \bibinfo {author} {\bibfnamefont {D.~S.}\ \bibnamefont {Grebenkov}}, \ and\
  \bibinfo {author} {\bibfnamefont {R.}~\bibnamefont {Metzler}},\ }\href@noop
  {} {\bibfield  {journal} {\bibinfo  {journal} {New Journal of Physics}\
  }\textbf {\bibinfo {volume} {18}},\ \bibinfo {pages} {013027} (\bibinfo
  {year} {2016})}\BibitemShut {NoStop}%
\bibitem [{\citenamefont {Cherstvy}\ and\ \citenamefont
  {Metzler}(2013)}]{Cherstvy2013}%
  \BibitemOpen
  \bibfield  {author} {\bibinfo {author} {\bibfnamefont {A.~G.}\ \bibnamefont
  {Cherstvy}}\ and\ \bibinfo {author} {\bibfnamefont {R.}~\bibnamefont
  {Metzler}},\ }\href@noop {} {\bibfield  {journal} {\bibinfo  {journal}
  {Physical Chemistry Chemical Physics}\ }\textbf {\bibinfo {volume} {15}},\
  \bibinfo {pages} {20220} (\bibinfo {year} {2013})}\BibitemShut {NoStop}%
\bibitem [{\citenamefont {Schulz}\ \emph {et~al.}(2013)\citenamefont {Schulz},
  \citenamefont {Barkai},\ and\ \citenamefont {Metzler}}]{Schulz2013}%
  \BibitemOpen
  \bibfield  {author} {\bibinfo {author} {\bibfnamefont {J.~H.~P.}\
  \bibnamefont {Schulz}}, \bibinfo {author} {\bibfnamefont {E.}~\bibnamefont
  {Barkai}}, \ and\ \bibinfo {author} {\bibfnamefont {R.}~\bibnamefont
  {Metzler}},\ }\href@noop {} {\bibfield  {journal} {\bibinfo  {journal}
  {Physical Review Letters}\ }\textbf {\bibinfo {volume} {110}},\ \bibinfo
  {pages} {020602} (\bibinfo {year} {2013})}\BibitemShut {NoStop}%
\bibitem [{\citenamefont {Schulz}\ \emph {et~al.}(2014)\citenamefont {Schulz},
  \citenamefont {Barkai},\ and\ \citenamefont {Metzler}}]{Schulz2014}%
  \BibitemOpen
  \bibfield  {author} {\bibinfo {author} {\bibfnamefont {J.~H.~P.}\
  \bibnamefont {Schulz}}, \bibinfo {author} {\bibfnamefont {E.}~\bibnamefont
  {Barkai}}, \ and\ \bibinfo {author} {\bibfnamefont {R.}~\bibnamefont
  {Metzler}},\ }\href@noop {} {\bibfield  {journal} {\bibinfo  {journal}
  {Physical Review X}\ }\textbf {\bibinfo {volume} {4}},\ \bibinfo {pages}
  {011028} (\bibinfo {year} {2014})}\BibitemShut {NoStop}%
\bibitem [{\citenamefont {Grebenkov}\ \emph {et~al.}(2018)\citenamefont
  {Grebenkov}, \citenamefont {Metzler},\ and\ \citenamefont
  {Oshanin}}]{Grebenkov2018}%
  \BibitemOpen
  \bibfield  {author} {\bibinfo {author} {\bibfnamefont {D.~S.}\ \bibnamefont
  {Grebenkov}}, \bibinfo {author} {\bibfnamefont {R.}~\bibnamefont {Metzler}},
  \ and\ \bibinfo {author} {\bibfnamefont {G.}~\bibnamefont {Oshanin}},\
  }\href@noop {} {\bibfield  {journal} {\bibinfo  {journal} {Communications
  Chemistry}\ }\textbf {\bibinfo {volume} {1}},\ \bibinfo {pages} {96}
  (\bibinfo {year} {2018})}\BibitemShut {NoStop}%
\bibitem [{\citenamefont {Tarasova}\ and\ \citenamefont
  {Nerukh}(2018)}]{Tarasova2018}%
  \BibitemOpen
  \bibfield  {author} {\bibinfo {author} {\bibfnamefont {E.}~\bibnamefont
  {Tarasova}}\ and\ \bibinfo {author} {\bibfnamefont {D.}~\bibnamefont
  {Nerukh}},\ }\href@noop {} {\bibfield  {journal} {\bibinfo  {journal}
  {Journal of Physical Chemistry Letters}\ }\textbf {\bibinfo {volume} {9}},\
  \bibinfo {pages} {5805} (\bibinfo {year} {2018})}\BibitemShut {NoStop}%
\bibitem [{\citenamefont {Hopkins}\ \emph {et~al.}(2010)\citenamefont
  {Hopkins}, \citenamefont {Fortini}, \citenamefont {Archer},\ and\
  \citenamefont {Schmidt}}]{Hopkins2010}%
  \BibitemOpen
  \bibfield  {author} {\bibinfo {author} {\bibfnamefont {P.}~\bibnamefont
  {Hopkins}}, \bibinfo {author} {\bibfnamefont {A.}~\bibnamefont {Fortini}},
  \bibinfo {author} {\bibfnamefont {A.~J.}\ \bibnamefont {Archer}}, \ and\
  \bibinfo {author} {\bibfnamefont {M.}~\bibnamefont {Schmidt}},\ }\href@noop
  {} {\bibfield  {journal} {\bibinfo  {journal} {Journal of Chemical Physics}\
  }\textbf {\bibinfo {volume} {133}},\ \bibinfo {pages} {224505} (\bibinfo
  {year} {2010})}\BibitemShut {NoStop}%
\bibitem [{\citenamefont {Dyre}(2018)}]{Dyre2018}%
  \BibitemOpen
  \bibfield  {author} {\bibinfo {author} {\bibfnamefont {J.~C.}\ \bibnamefont
  {Dyre}},\ }\href@noop {} {\bibfield  {journal} {\bibinfo  {journal} {Journal
  of Chemical Physics}\ }\textbf {\bibinfo {volume} {149}},\ \bibinfo {pages}
  {210901} (\bibinfo {year} {2018})}\BibitemShut {NoStop}%
\bibitem [{\citenamefont {Goncharov}\ \emph {et~al.}(2013)\citenamefont
  {Goncharov}, \citenamefont {Melent'ev},\ and\ \citenamefont
  {Postnikov}}]{Goncharov2013}%
  \BibitemOpen
  \bibfield  {author} {\bibinfo {author} {\bibfnamefont {A.~L.}\ \bibnamefont
  {Goncharov}}, \bibinfo {author} {\bibfnamefont {V.~V.}\ \bibnamefont
  {Melent'ev}}, \ and\ \bibinfo {author} {\bibfnamefont {E.~B.}\ \bibnamefont
  {Postnikov}},\ }\href@noop {} {\bibfield  {journal} {\bibinfo  {journal}
  {European Physical Journal B}\ }\textbf {\bibinfo {volume} {86}},\ \bibinfo
  {pages} {357} (\bibinfo {year} {2013})}\BibitemShut {NoStop}%
\bibitem [{\citenamefont {Chor{\c{a}}{\.z}ewski}\ \emph
  {et~al.}(2017)\citenamefont {Chor{\c{a}}{\.z}ewski}, \citenamefont
  {Postnikov}, \citenamefont {Jasiok}, \citenamefont {Nedyalkov},\ and\
  \citenamefont {Jacquemin}}]{Chorazewski2017}%
  \BibitemOpen
  \bibfield  {author} {\bibinfo {author} {\bibfnamefont {M.}~\bibnamefont
  {Chor{\c{a}}{\.z}ewski}}, \bibinfo {author} {\bibfnamefont {E.~B.}\
  \bibnamefont {Postnikov}}, \bibinfo {author} {\bibfnamefont {B.}~\bibnamefont
  {Jasiok}}, \bibinfo {author} {\bibfnamefont {Y.~V.}\ \bibnamefont
  {Nedyalkov}}, \ and\ \bibinfo {author} {\bibfnamefont {J.}~\bibnamefont
  {Jacquemin}},\ }\href@noop {} {\bibfield  {journal} {\bibinfo  {journal}
  {Scientific Reports}\ }\textbf {\bibinfo {volume} {7}},\ \bibinfo {pages}
  {5563} (\bibinfo {year} {2017})}\BibitemShut {NoStop}%
\bibitem [{NIS()}]{NIST}%
  \BibitemOpen
  \href@noop {} {}\bibinfo {note}
  {\url{http://webbook.nist.gov/chemistry/fluid}}\BibitemShut {NoStop}%
\bibitem [{\citenamefont {Tegeler}\ \emph {et~al.}(1999)\citenamefont
  {Tegeler}, \citenamefont {Span},\ and\ \citenamefont {Wagner}}]{Tegeler1999}%
  \BibitemOpen
  \bibfield  {author} {\bibinfo {author} {\bibfnamefont {C.}~\bibnamefont
  {Tegeler}}, \bibinfo {author} {\bibfnamefont {R.}~\bibnamefont {Span}}, \
  and\ \bibinfo {author} {\bibfnamefont {W.}~\bibnamefont {Wagner}},\
  }\href@noop {} {\bibfield  {journal} {\bibinfo  {journal} {Journal of
  Physical and Chemical Reference Data}\ }\textbf {\bibinfo {volume} {28}},\
  \bibinfo {pages} {779} (\bibinfo {year} {1999})}\BibitemShut {NoStop}%
\bibitem [{git()}]{gitArgon}%
  \BibitemOpen
  \href@noop {} {}\bibinfo {note}
  {\url{https://github.com/KenNewcomb/LJ-Argon}}\BibitemShut {NoStop}%
\bibitem [{\citenamefont {Wei-Zhong}\ \emph {et~al.}(2008)\citenamefont
  {Wei-Zhong}, \citenamefont {Cong},\ and\ \citenamefont {Jian}}]{Wei2008}%
  \BibitemOpen
  \bibfield  {author} {\bibinfo {author} {\bibfnamefont {L.}~\bibnamefont
  {Wei-Zhong}}, \bibinfo {author} {\bibfnamefont {C.}~\bibnamefont {Cong}}, \
  and\ \bibinfo {author} {\bibfnamefont {Y.}~\bibnamefont {Jian}},\ }\href@noop
  {} {\bibfield  {journal} {\bibinfo  {journal} {Heat Transfer -- Asian
  Research}\ }\textbf {\bibinfo {volume} {37}},\ \bibinfo {pages} {86}
  (\bibinfo {year} {2008})}\BibitemShut {NoStop}%
\bibitem [{\citenamefont {Cini-Castagnoli}\ and\ \citenamefont
  {Ricci}(1960)}]{Cini1960}%
  \BibitemOpen
  \bibfield  {author} {\bibinfo {author} {\bibfnamefont {G.}~\bibnamefont
  {Cini-Castagnoli}}\ and\ \bibinfo {author} {\bibfnamefont {F.~P.}\
  \bibnamefont {Ricci}},\ }\href@noop {} {\bibfield  {journal} {\bibinfo
  {journal} {Il Nuovo Cimento}\ }\textbf {\bibinfo {volume} {15}},\ \bibinfo
  {pages} {795} (\bibinfo {year} {1960})}\BibitemShut {NoStop}%
\bibitem [{\citenamefont {Naghizadeh}\ and\ \citenamefont
  {Rice}(1962)}]{Naghizadeh1962}%
  \BibitemOpen
  \bibfield  {author} {\bibinfo {author} {\bibfnamefont {J.}~\bibnamefont
  {Naghizadeh}}\ and\ \bibinfo {author} {\bibfnamefont {S.~A.}\ \bibnamefont
  {Rice}},\ }\href@noop {} {\bibfield  {journal} {\bibinfo  {journal} {Journal
  of Chemical Physics}\ }\textbf {\bibinfo {volume} {36}},\ \bibinfo {pages}
  {2710} (\bibinfo {year} {1962})}\BibitemShut {NoStop}%
\bibitem [{\citenamefont {Bodrova}\ \emph {et~al.}(2016)\citenamefont
  {Bodrova}, \citenamefont {Chechkin}, \citenamefont {Cherstvy}, \citenamefont
  {Safdari}, \citenamefont {Sokolov},\ and\ \citenamefont
  {Metzler}}]{Bodrova2016}%
  \BibitemOpen
  \bibfield  {author} {\bibinfo {author} {\bibfnamefont {A.~S.}\ \bibnamefont
  {Bodrova}}, \bibinfo {author} {\bibfnamefont {A.~V.}\ \bibnamefont
  {Chechkin}}, \bibinfo {author} {\bibfnamefont {A.~G.}\ \bibnamefont
  {Cherstvy}}, \bibinfo {author} {\bibfnamefont {H.}~\bibnamefont {Safdari}},
  \bibinfo {author} {\bibfnamefont {I.~M.}\ \bibnamefont {Sokolov}}, \ and\
  \bibinfo {author} {\bibfnamefont {R.}~\bibnamefont {Metzler}},\ }\href@noop
  {} {\bibfield  {journal} {\bibinfo  {journal} {Scientific reports}\ }\textbf
  {\bibinfo {volume} {6}},\ \bibinfo {pages} {30520} (\bibinfo {year}
  {2016})}\BibitemShut {NoStop}%
\bibitem [{\citenamefont {Fa}(2018)}]{Fa2018book}%
  \BibitemOpen
  \bibfield  {author} {\bibinfo {author} {\bibfnamefont {K.~S.}\ \bibnamefont
  {Fa}},\ }\href@noop {} {\emph {\bibinfo {title} {{Langevin and Fokker-Planck
  Equations and Their Generalizations: Descriptions and Solutions}}}}\
  (\bibinfo  {publisher} {World Scientific},\ \bibinfo {year}
  {2018})\BibitemShut {NoStop}%
\bibitem [{\citenamefont {Filip}\ \emph {et~al.}(2019)\citenamefont {Filip},
  \citenamefont {Javeed},\ and\ \citenamefont {Trefethen}}]{Filip2019}%
  \BibitemOpen
  \bibfield  {author} {\bibinfo {author} {\bibfnamefont {S.}~\bibnamefont
  {Filip}}, \bibinfo {author} {\bibfnamefont {A.}~\bibnamefont {Javeed}}, \
  and\ \bibinfo {author} {\bibfnamefont {L.~N.}\ \bibnamefont {Trefethen}},\
  }\href@noop {} {\bibfield  {journal} {\bibinfo  {journal} {SIAM Review}\
  }\textbf {\bibinfo {volume} {61}},\ \bibinfo {pages} {185} (\bibinfo {year}
  {2019})}\BibitemShut {NoStop}%
\bibitem [{\citenamefont {Rosenfeld}(1977)}]{Rosenfeld1977}%
  \BibitemOpen
  \bibfield  {author} {\bibinfo {author} {\bibfnamefont {Y.}~\bibnamefont
  {Rosenfeld}},\ }\href@noop {} {\bibfield  {journal} {\bibinfo  {journal}
  {Physical Review A}\ }\textbf {\bibinfo {volume} {15}},\ \bibinfo {pages}
  {2545} (\bibinfo {year} {1977})}\BibitemShut {NoStop}%
\bibitem [{\citenamefont {Dzugutov}(1996)}]{Dzugutov1996}%
  \BibitemOpen
  \bibfield  {author} {\bibinfo {author} {\bibfnamefont {M.}~\bibnamefont
  {Dzugutov}},\ }\href@noop {} {\bibfield  {journal} {\bibinfo  {journal}
  {Nature}\ }\textbf {\bibinfo {volume} {381}},\ \bibinfo {pages} {137}
  (\bibinfo {year} {1996})}\BibitemShut {NoStop}%
\bibitem [{\citenamefont {Rosenfeld}(1999)}]{Rosenfeld1999}%
  \BibitemOpen
  \bibfield  {author} {\bibinfo {author} {\bibfnamefont {Y.}~\bibnamefont
  {Rosenfeld}},\ }\href@noop {} {\bibfield  {journal} {\bibinfo  {journal}
  {Journal of Physics: Condensed Matter}\ }\textbf {\bibinfo {volume} {11}},\
  \bibinfo {pages} {5415} (\bibinfo {year} {1999})}\BibitemShut {NoStop}%
\bibitem [{\citenamefont {Baranyai}\ and\ \citenamefont
  {Evans}(1989)}]{Baranyai1989}%
  \BibitemOpen
  \bibfield  {author} {\bibinfo {author} {\bibfnamefont {A.}~\bibnamefont
  {Baranyai}}\ and\ \bibinfo {author} {\bibfnamefont {D.~J.}\ \bibnamefont
  {Evans}},\ }\href@noop {} {\bibfield  {journal} {\bibinfo  {journal}
  {Physical Review A}\ }\textbf {\bibinfo {volume} {40}},\ \bibinfo {pages}
  {3817} (\bibinfo {year} {1989})}\BibitemShut {NoStop}%
\bibitem [{\citenamefont {Saija}\ and\ \citenamefont
  {Giaquinta}(1996)}]{Saija1996}%
  \BibitemOpen
  \bibfield  {author} {\bibinfo {author} {\bibfnamefont {F.}~\bibnamefont
  {Saija}}\ and\ \bibinfo {author} {\bibfnamefont {P.~V.}\ \bibnamefont
  {Giaquinta}},\ }\href@noop {} {\bibfield  {journal} {\bibinfo  {journal}
  {Journal of Physics: Condensed Matter}\textbf {\bibinfo {volume} {8}}\ ,\ \bibinfo {pages} {8137}} (\bibinfo
  {year} {1996})}\BibitemShut {NoStop}%
\bibitem [{\citenamefont {Vaz}\ \emph {et~al.}(2012)\citenamefont {Vaz},
  \citenamefont {Magalh\~aes}, \citenamefont {Fernandes},\ and\ \citenamefont
  {Silva}}]{Vaz2012}%
  \BibitemOpen
  \bibfield  {author} {\bibinfo {author} {\bibfnamefont {R.~V.}\ \bibnamefont
  {Vaz}}, \bibinfo {author} {\bibfnamefont {A.~L.}\ \bibnamefont
  {Magalh\~aes}}, \bibinfo {author} {\bibfnamefont {D.~L.}\ \bibnamefont
  {Fernandes}}, \ and\ \bibinfo {author} {\bibfnamefont {C.~M.}\ \bibnamefont
  {Silva}},\ }\href@noop {} {\bibfield  {journal} {\bibinfo  {journal}
  {Chemical Engineering Science}\ }\textbf {\bibinfo {volume} {79}},\ \bibinfo
  {pages} {153} (\bibinfo {year} {2012})}\BibitemShut {NoStop}%
\bibitem [{\citenamefont {J{\o}rgensen}(1997)}]{Jorgensen1997book}%
  \BibitemOpen
  \bibfield  {author} {\bibinfo {author} {\bibfnamefont {B.}~\bibnamefont
  {J{\o}rgensen}},\ }\href@noop {} {\emph {\bibinfo {title} {The theory of
  dispersion models}}}\ (\bibinfo  {publisher} {Chapman and Hall, London},\
  \bibinfo {year} {1997})\BibitemShut {NoStop}%
\bibitem [{\citenamefont {Landau}\ and\ \citenamefont
  {M.}(2013)}]{Landau2013statphys}%
  \BibitemOpen
  \bibfield  {author} {\bibinfo {author} {\bibfnamefont {L.~D.}\ \bibnamefont
  {Landau}}\ and\ \bibinfo {author} {\bibfnamefont {L.~E.}\ \bibnamefont
  {M.}},\ }\href@noop {} {\emph {\bibinfo {title} {Statistical Physics. Part
  1.}}}\ (\bibinfo  {publisher} {Elsevier},\ \bibinfo {year}
  {2013})\BibitemShut {NoStop}%
\bibitem [{\citenamefont {v~Smoluchowski}(1904)}]{Smoluchowski1904}%
  \BibitemOpen
  \bibfield  {author} {\bibinfo {author} {\bibfnamefont {M.}~\bibnamefont
  {v~Smoluchowski}},\ }in\ \href@noop {} {\emph {\bibinfo {booktitle}
  {Festschrift Ludwig Boltzmann gewidmet zum sechzigsten Geburtstage 20.
  Februar 1904}}}\ (\bibinfo  {publisher} {J.A. Barth, Leipzig},\ \bibinfo
  {year} {1904})\BibitemShut {NoStop}%
\bibitem [{\citenamefont {Thiel}\ \emph {et~al.}(2016)\citenamefont {Thiel},
  \citenamefont {Sokolov},\ and\ \citenamefont {Postnikov}}]{Thiel2016}%
  \BibitemOpen
  \bibfield  {author} {\bibinfo {author} {\bibfnamefont {F.}~\bibnamefont
  {Thiel}}, \bibinfo {author} {\bibfnamefont {I.~M.}\ \bibnamefont {Sokolov}},
  \ and\ \bibinfo {author} {\bibfnamefont {E.~B.}\ \bibnamefont {Postnikov}},\
  }\href@noop {} {\bibfield  {journal} {\bibinfo  {journal} {Physical Review
  E}\ }\textbf {\bibinfo {volume} {93}},\ \bibinfo {pages} {052104} (\bibinfo
  {year} {2016})}\BibitemShut {NoStop}%
\bibitem [{\citenamefont {Cherstvy}\ \emph {et~al.}(2018)\citenamefont
  {Cherstvy}, \citenamefont {Thapa}, \citenamefont {Mardoukhi}, \citenamefont
  {Chechkin},\ and\ \citenamefont {Metzler}}]{Cherstvy2018}%
  \BibitemOpen
  \bibfield  {author} {\bibinfo {author} {\bibfnamefont {A.~G.}\ \bibnamefont
  {Cherstvy}}, \bibinfo {author} {\bibfnamefont {S.}~\bibnamefont {Thapa}},
  \bibinfo {author} {\bibfnamefont {Y.}~\bibnamefont {Mardoukhi}}, \bibinfo
  {author} {\bibfnamefont {A.~V.}\ \bibnamefont {Chechkin}}, \ and\ \bibinfo
  {author} {\bibfnamefont {R.}~\bibnamefont {Metzler}},\ }\href@noop {}
  {\bibfield  {journal} {\bibinfo  {journal} {Physical Review E}\ }\textbf
  {\bibinfo {volume} {98}},\ \bibinfo {pages} {022134} (\bibinfo {year}
  {2018})}\BibitemShut {NoStop}%
\bibitem [{\citenamefont {Ku{\'s}mierz}\ \emph {et~al.}(2018)\citenamefont
  {Ku{\'s}mierz}, \citenamefont {Dybiec},\ and\ \citenamefont
  {Gudowska-Nowak}}]{Kusmierz2018}%
  \BibitemOpen
  \bibfield  {author} {\bibinfo {author} {\bibfnamefont {L.}~\bibnamefont
  {Ku{\'s}mierz}}, \bibinfo {author} {\bibfnamefont {B.}~\bibnamefont
  {Dybiec}}, \ and\ \bibinfo {author} {\bibfnamefont {E.}~\bibnamefont
  {Gudowska-Nowak}},\ }\href@noop {} {\bibfield  {journal} {\bibinfo  {journal}
  {Entropy}\ }\textbf {\bibinfo {volume} {20}},\ \bibinfo {pages} {658}
  (\bibinfo {year} {2018})}\BibitemShut {NoStop}%
\bibitem [{\citenamefont {Pellicane}(2012)}]{Pellicane2012}%
  \BibitemOpen
  \bibfield  {author} {\bibinfo {author} {\bibfnamefont {G.}~\bibnamefont
  {Pellicane}},\ }\href@noop {} {\bibfield  {journal} {\bibinfo  {journal}
  {Journal of Physical Chemistry B}\ }\textbf {\bibinfo {volume} {116}},\
  \bibinfo {pages} {2114} (\bibinfo {year} {2012})}\BibitemShut {NoStop}%
\bibitem [{pol()}]{polyfit}%
  \BibitemOpen
  \href@noop {} {}\bibinfo {note}
  {\url{https://www.mathworks.com/help/matlab/ref/polyfit.html}}\BibitemShut
  {NoStop}%
\end{thebibliography}

%

\end{document}